\documentclass[aps, twocolumn, superscriptaddress, floatfix]{revtex4-1}

\usepackage{graphicx, color}
\usepackage{amsmath, amssymb}
\usepackage[colorlinks = true,
			citecolor = blue,
			linkcolor = blue]{hyperref}
\usepackage[normalem]{ulem}

\newcommand{\Iob}{I_{\rm ob}}

\newcommand{\If}{I_{\rm f}}
\newcommand{\psif}{\psi_{\rm f}}
\newcommand{\kf}{k_{\rm f}}

\newcommand{\cs}{c_{\rm s}}

\newcommand{\nr}{n_{\rm e}}
\newcommand{\ttt}[1]{\times 10^{#1}}

\begin{document}

\title{Experimental observation of turbulent coherent structures in a superfluid of light}

\author{A. Eloy}
\affiliation{Universit\'e C\^ote d'Azur, CNRS, INPHYNI, France}
\author{O. Boughdad}
\affiliation{Universit\'e C\^ote d'Azur, CNRS, INPHYNI, France}
\author{M. Albert}
\affiliation{Universit\'e C\^ote d'Azur, CNRS, INPHYNI, France}
\author{P.-\'E. Larr\'e}
\affiliation{Universit\'e C\^ote d'Azur, CNRS, INPHYNI, France}
\author{F. Mortessagne}
\affiliation{Universit\'e C\^ote d'Azur, CNRS, INPHYNI, France}
\author{M. Bellec}
\email{matthieu.bellec@inphyni.cnrs.fr}
\affiliation{Universit\'e C\^ote d'Azur, CNRS, INPHYNI, France}
\author{C. Michel}
\email{claire.michel@inphyni.cnrs.fr}
\affiliation{Universit\'e C\^ote d'Azur, CNRS, INPHYNI, France}

\begin{abstract}
We experimentally explore the rich variety of nonlinear coherent structures arising in a turbulent flow of superfluid light past an obstacle in an all-optical configuration. The different hydrodynamic regimes observed are organised in a unique phase diagram involving the velocity of the flow and the diameter of the obstacle. Then, we focus on the vortices nucleated in the wake of the obstacle by investigating their intensity profile and the dependence of the radius of their core on the healing length. Our results pave the way for further investigations on turbulence in photon superfluids and provide versatile experimental tools for simulating quantum transport with nonlinear light.
\end{abstract}

\maketitle

\maketitle

\begin{figure*}[t]
\centering
\includegraphics{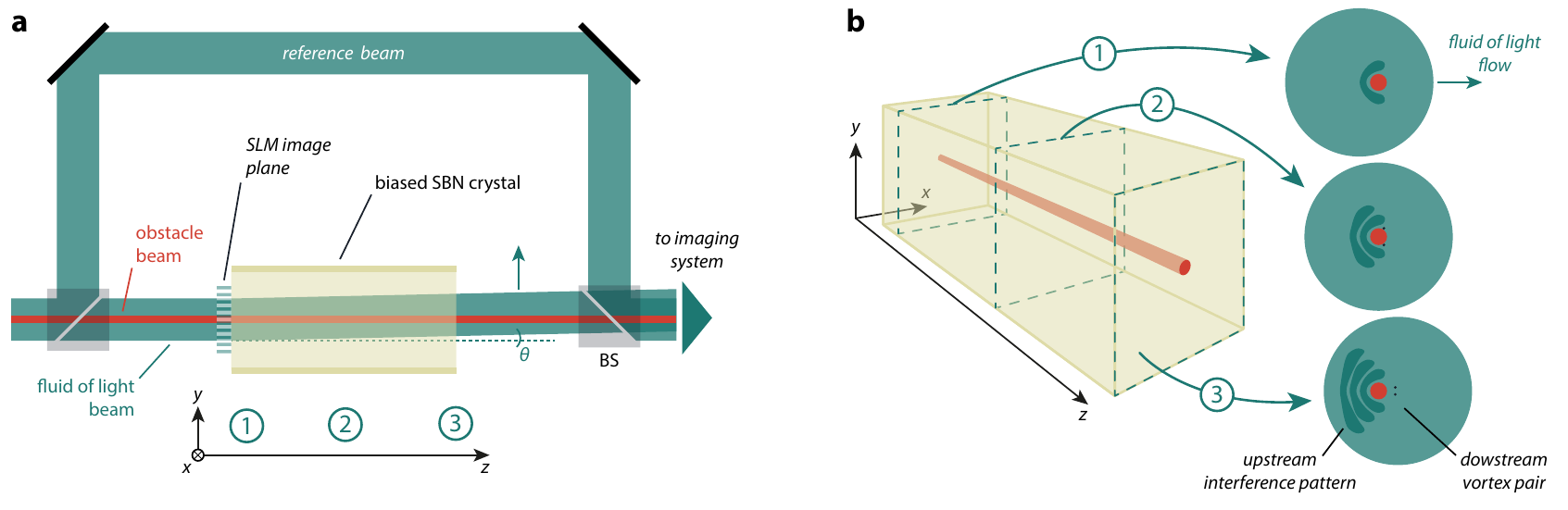}
\caption{(a) Sketch of the experimental setup. The narrow red beam creates a $z$-invariant localised optical defect which acts as a potential barrier in the transverse plane $(x,y)$. The large green beam creates a 2D fluid of light evolving along the $z$-axis. The input angle $\theta$ of the green beam with respect to the red beam is tuned via a spatial light modulator (SLM) conjugated with a digital micro-mirror device (DMD). It controls the transverse velocity of the fluid of light (indicated by the green arrow). Both are propagating through a biased nonlinear SBN crystal and are imaged via a microscope objective on a sCMOS camera. Beam splitters (BS) are used to get a reference beam to build the interferogram and recover both the amplitude and the phase of the fluid of light.
(b) Three-step sketch of the evolution of the 2D fluid of light (green) past the obstacle (red). The propagation coordinate $z$ plays the role of time. The third step shows the formation of a typical interference pattern upstream and a vortex-pair generation downstream from the obstacle.}
\label{fig:fig_1}
\end{figure*}

\section{Introduction}

Turbulence appears in ordinary fluids as simple as air or water, but also in astrophysical systems \cite{brandenburg2011}, plasmas \cite{yoshizawa2003} and superfluids \cite{skrbek2011, kolmakov2013}.
The main features shared by the different types of turbulence are the unpredictable and irregular character of the system dynamics far from equilibrium, and the interplay between many different physical length scales through highly nonlinear processes \cite{Sreenivasan1999}. 
Quantum turbulence mainly differs from turbulence in ordinary fluids because, in quantum fluids, vorticity is quantised and constrained to appear at a single characteristic length scale \cite{pitaevskii2003}.
Examples of quantum fluids displaying quantum turbulence are superfluid helium \cite{allen38, kapitsa38}, atomic Bose-Einstein condensates \cite{pitaevskii2003, neely2010, bloch2012} and quantum fluids of light \cite{carusotto13}.\\
Merging nonlinear optics and quantum hydrodynamics, quantum fluids of light have gained great interest in the past few years. Indeed, in properly engineered experimental optical devices, photons can acquire an effective mass and be in a fully controlled effective interaction. 
They behave collectively as a quantum fluid, and share remarkable common features with other systems such as superfluidity and quantum turbulence. Quantum fluids of light have been investigated in one, two and three effective spatial dimensions (1, 2 and 3D) in various photonic platforms. 
Among the latter, the major ones are semiconductor \cite{deng2010, amo2009, amo2011, nardin2011, sanvitto2011, carusotto13, pigeon2011, pigeon2017, claude2020, lerario2020} and optical \cite{kurtscheid2019, chiao1999, chiao2000} microcavities, and cavityless bulk optical media \cite{frisch1992, pomeau1993, leboeuf2010, carusotto2014, larre2012, larre2015, larre2015bis} where the light propagation axis plays the role of time, as sketched in fig. \ref{fig:fig_1}.\\
In the latter propagating geometry (and in a regime where quantum fluctuations \cite{larre2015bis} are negligible), fluids of light have been produced in several nonlinear media ranging from liquid crystals \cite{bortolozzo09, laurie2012} to thermal liquids \cite{vocke2015, vocke2016, vocke2018} and atomic vapours \cite{santic2018, fontaine2018, fontaine2020}, but also photorefractive (PR) crystals \cite{wan2007, wan2008, wan2010, sun2012, situ2020} in which a quantitative measurement of the normal/superfluid transition has been recently performed \cite{michel2018}.
These optical systems allow for a full control and flexibility in the generation, manipulation and observation of photon fluids in various dimensions.
In addition, the nonlinear interactions and the external potentials experienced by the fluid can be precisely customised. 
Consequently, they consist in versatile systems, fully designated to investigate light hydrodynamics in different kinds of environments, from the simplest case of a homogeneous medium \cite{fontaine2018, fontaine2020} to more complex landscapes with a single obstacle \cite{vocke2016, michel2018} or spatial disorder.\\
One of the simplest way to generate turbulence in these systems is to insert an obstacle in the fluid flow. By accurately tuning its velocity, the fluid enters a turbulent regime, giving birth to nonlinear coherent structures ranging from vortex pairs to dark solitons and snake instabilities \cite{armaroli2009, gallemi2016, lerario2020, claude2020}.\\
Although optical vortices have been widely studied within the last three decades \cite{coullet1989, shen2019}, their generation through the process of turbulence have been subjected to few studies, either numerical \cite{frisch1992, derossi1997, coddington2004, armaroli2009, manjun2010, aioi2011} or experimental \cite{mamaev1996b, raman2001, gorza2006, lagoudakis2008, kwon2015, lerario2020, claude2020}. Related works on the Berezinskii-Kosterlitz-Thouless transition and its underlying vortex dynamics have been recently brought to the photonic realm \cite{small2011, roumpos2012, caputo2018, situ2020, comaron2020}.

In this letter, we propose a systematic experimental study of the different turbulent coherent structures generated in the wake of a 2D fluid of light passing an obstacle in a propagating geometry. We gather our experimental data in a diagram within regions of apparition of the different nonlinear hydrodynamic structures as a function of the fluid velocity and the obstacle diameter, which is not the standard parameter used in related studies \cite{hakim97,leboeuf2001,pavloff2002,larre2012,larre2015}. We dedicate the last section to a detailed investigation of the intensity profiles of isolated vortices and of the dependence of their radius on the healing length.

\section{Fluid reformulation of light propagation}

The propagation in the positive-$z$ direction of a continuous laser beam along a nonlinear PR crystal with a $z$-invariant optical obstacle is ruled in the scalar, paraxial and monochromatic approximations by the following 2D nonlinear Schr\"odinger equation for the complex envelope $\psif(\mathbf{r}=(x,y),z)$ of the optical field \cite{boyd2008}:
\begin{equation}
    i\partial_z \psif = -\frac{1}{2\nr\kf}\nabla_{\mathbf{r}}^2 \psif - \kf\Delta n(|\psif|^2)\psif - \kf\delta n(\mathbf{r}) \psif.
	\label{eq:schro}
\end{equation}
The different quantities involved in this equation are defined in the following lines.
The analogy with the genuine Gross-Pitaevskii equation for the wavefunction of a quasi-2D weakly interacting Bose-Einstein condensate \cite{pitaevskii2003} is a well known fact \cite{frisch1992, pomeau1993, chiao1999, chiao2000}. Importantly, the behaviour of the laser beam in the transverse $\mathbf{r}$ plane mimics the flow of such a matter fluid past an obstacle, the propagation coordinate $z$ playing the role of time, thus defining a \textsl{fluid of light} \cite{carusotto2014}.
In this context, the propagation constant $\nr\kf=2\pi\nr/\lambda_{\rm f}$ of the beam of wavelength $\lambda_{\rm f}$ in the medium of refractive index $\nr$ is the optical analog of the atom mass. 
On the other hand, $\Delta n(|\psif|^2)<0$, responsible for the defocusing nonlinear response of the crystal, mimics repulsive photon-photon interactions and acts over the whole transverse extension of the fluid of light. 
In this quantum hydrodynamic description of the laser beam, the optical intensity $\If=|\psif|^{2}$ corresponds to the local density of the fluid of light and the gradient of the optical phase $\arg(\psif)$ gives the local flow velocity $\mathbf{v}=\nabla_{\mathbf{r}}\arg(\psif)/(\nr\kf)$.
In the paraxial limit, a uniform initial velocity is given by the uniform phase of a plane wave such as $v\simeq\theta/\nr$, with $\theta$ being the angle between the beam and the $z$ axis (green beam of the bottom arm in fig.~\ref{fig:fig_1}(a)). 
Finally, in the last term of eq.~(\ref{eq:schro}), $\delta n(\mathbf{r})$ is a $z$-invariant refractive-index depletion which is photo-induced by a second laser beam (red beam in fig. \ref{fig:fig_1}(a)).
This emulates the presence of a penetrable obstacle for the fluid of light. The height and the size of such a potential barrier depends on the intensity $\Iob$ and the diameter $d$ of this secondary laser beam \cite{michel2018}.
We also define, away from the obstacle where the fluid of light remains unperturbed, an analog healing length $\xi=[\nr\kf^2|\Delta n(\If)|]^{-1/2}$, which corresponds to the smallest length scale for density modulations in the transverse plane, and an analog sound velocity $\cs=(\nr\kf\xi)^{-1}=[|\Delta n(\If)|/\nr]^{1/2}$ for the modulation waves of the fluid of light. Both analog quantities depend on the fluid density, and are then experimentally controlled via the intensity $\If$.\\
Panels 1, 2 and 3 in fig. \ref{fig:fig_1}(b) correspond to three typical density snapshots of the evolution (along the $z$ axis) of the fluid of light (green spots), flowing from left to right (along the $x$ axis) around the obstacle (red spots). The oscillatory pattern and the density drops illustrated in the third panel are particularly representative. The former, reminiscent of interferences between the incident and reflected components of the green beam onto the red one, corresponds, in the hydrodynamics language, to linear or cnoidal periodic excitations radiated far away from the obstacle \cite{khamis2008}. The latter is representative of a regime of turbulent flow characterised by the nucleation of vortex pairs \cite{pomeau1993}, which one studies in this letter. 

\begin{figure*}[t]
\centering
\includegraphics{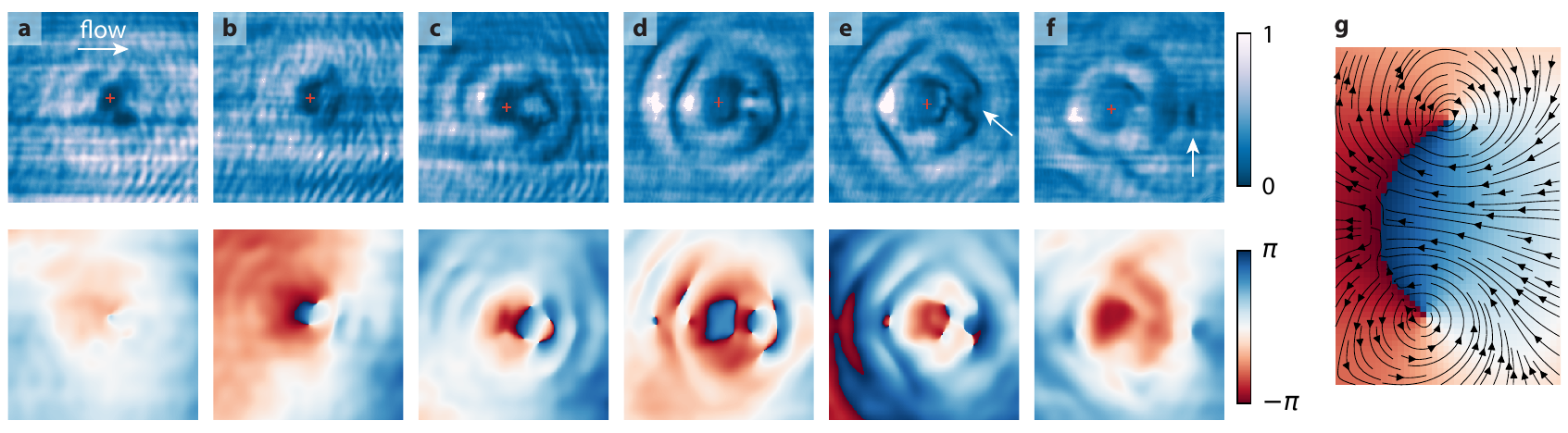}
\caption{(a--f) Images of the intensity (top row) and of the phase (bottom row) of the 2D fluid of light at the output of the nonlinear crystal for various values of the Mach number $v/c_{\rm s}$ and of the obstacle diameter $d$. The fluid of light flows from left to right and the central position of the diameter-varying obstacle is indicated by the red crosses. Each image is 245 $\mu$m$\times$245 $\mu$m. The intensity images are normalised to their maximum value. (a) Superfluid phase, $d$ = 50 $\mu$m. (b--d) Regime of vortex-pair generation. $d$ = 55, 65, 80 $\mu$m for (b), (c) and (d), respectively. (e) Regime of snake instability, indicated by the white arrow. $d$ = 80 $\mu$m.  (f) Solitonic regime. A dark soliton is indicated by the white arrow. $d$ = 90 $\mu$m. No phase singularities are associated to the structures observed in (e) and (f).
(g) Zoom of the phase image (d, bottom panel). The region of interest corresponds to the two clear localised intensity drops observed in the intensity image (d, top panel). The phase derivative (\textit{i.e.}, the fluid of light velocity field, see the text) stream plot (black lines and arrows) indicates that the vortices of the pair have opposite circulations.}
\label{fig:fig_2}
\end{figure*}

\section{Experimental implementation}

In the experiments, the nonlinear medium considered is a $5\times5\times20$ mm$^{3}$ strontium barium niobate (SBN:61) photorefractive crystal additionally doped with cerium (0.002\%) to enhance its photoconductivity.
Because of the saturable nature of the PR effect, we take care of always working in a Kerr-like regime ($\Delta n(|\psif|^2)\propto|\psif|^2$)
where the correspondence between eq. \eqref{eq:schro} and the Gross-Pitaevskii equation is strict (see Supplementary Material for details).\\
The fluid of light laser beam is a cw source delivering a collimated Gaussian beam with a $460$ $\mu$m $1/e^2$ radius at $\lambda_{\rm f}$ = 532 nm. The input angle $\theta$ is tuned from 0 to $\pm 23$ mrad by means of a spatial light modulator (SLM) conjugated with a digital micro-mirror device (DMD) optical system.\\
The same conjugated devices are used to create the $z$-invariant obstacle which consists in a diffraction-free Bessel cw laser beam operating at $\lambda_{\rm ob}$ = 633 nm. The  diameter $d$ of the obstacle (defined as the diameter of the first zero of the rotation-invariant Bessel distribution) can be adjusted from 10 to 90 $\mu$m.
Both laser beams are linearly-polarised along the extraordinary axis to maximise the PR effect. The photo-induced refractive-index modification is of the order of $-10^{-4}$ (see Supplementary Material for  details).\\
For the detection part, an imaging system composed of a $\times$20 microscope objective ($\mathrm{NA}=0.4$) and of a sCMOS camera allows to get the spatial distribution of the near-field intensity of the laser beams at the output of the crystal. 
The interference between the fluid of light beam and a reference beam, recombined before the camera, allows to obtain an interferogram (the fringes are separated by a $2\pi$ spatial phase shift). Using the off-axis holography method \cite{leith1962, verrier2011}, both the amplitude and the phase of the optical fields can be reconstructed.

To summarise, our experimental apparatus makes it possible to i) prepare the initial condition of the 2D fluid of light, ii) design an obstacle with controllable strength and size, and iii) measure the hydrodynamic parameters of the fluid of light, \textit{i.e.}, both its density and velocity, at the output of the nonlinear crystal.

\section{Diagram of existence of the turbulent coherent structures in the fluid}

The Mach number, defined as $M=v/\cs$, is a crucial quantity that allows to run through all the different hydrodynamic regimes of the fluid of light. In our optical setup, $M$ typically varies from 0 to 3. As shown in \cite{michel2018}, for an obstacle diameter of the order of $\xi$, a crossover from a normal (\textit{i.e.}, nonsuperfluid but stationary) to a superfluid regime occurs at $M \sim 0.4$. This deviation from the Landau criterion \cite{leggett1999} is actually not surprising. Indeed, this criterion, which predicts a sharp normal/superfluid transition at $M=1$, is intrinsically valid for small obstacles and conservative dynamics. Therefore, it cannot be strictly applied to our setup where the optical defect is not of weak amplitude and where photon absorption is non-negligible (about 1.5 dB/cm).\\
Beyond Landau's theory, the critical velocity for superfluidity is smaller than the speed of sound and a nontrivial function of the obstacle parameters  (see, \textit{e.g.}, refs. \cite{hakim97, leboeuf2001, pavloff2002, larre2012, larre2015, albert2010, Engels2007, Hulet2010} for related studies in 1D), and the very nature of the nonsuperfluid/superfluid transition is modified by the appearance of an intermediate nonstationary turbulent regime. The latter is characterised by the advent of nonlinear coherent structures such as dark solitons undergoing a snake instability before breaking into pairs of quantised vortices of opposite circulations \cite{brand2002}. All these structures obviously depend on photon absorption in the present setup while superfluid features have been shown to be robust against dissipation \cite{wouters2010}.

The aim of this section is to measure such a phase diagram by systematically varying the Mach number of the flow, $v/\cs$, and the obstacle diameter normalised to the healing length, $d/\xi$, which is here the parameter used to
vary the state of the obstacle, as explained below.
Most theoretical studies build a diagram of existence of the different hydrodynamic coherent structures emerging in the fluid by varying the Mach number, as we do, and the height of the obstacle potential \cite{hakim97, leboeuf2001, pavloff2002, albert2008, larre2012, larre2015}. In our experimental apparatus, the latter variation would be performed by changing the depth of the refractive-index modulation $\delta n(\mathbf{r})$ induced by the obstacle beam. However, the range of accessible values is rather limited. Moreover, changing the value of the optical index imposes to adapt in a synchronised manner the intensities of both the fluid of light and obstacle beams (see Supplementary Material for details). But changing the intensity of the fluid of light beam would impact the values of $\xi$ and $\cs$, and thus dramatically change the global state of the system.
In this study we then focus on varying another intrinsic parameter of the obstacle, namely, its diameter $d$, which has the advantage of not altering the state of the incident fluid. This configuration has also been studied theoretically \cite{huepe2000, pinsker2014}, but has, to our knowledge, never been explored experimentally.

The fluid of light and obstacle beam intensities are respectively fixed to $\If=42~\mathrm{mW}/\mathrm{cm}^2$ and $\Iob=4P_{\rm ob}/(\pi d^2)=510~\mathrm{mW}/\mathrm{cm}^2$ at $z=0$, where $P_{\rm ob}$ is the input power of the obstacle beam. $\If$ above provides constant values for $\xi$ and $\cs$ while $\Iob$ is kept constant by simultaneously varying $P_{\rm ob}$ and $d$. The refractive index depletion corresponding to the obstacle is then fixed at a value $\delta n = -1.7\ttt{-4} = 0.8\,\delta n_{\rm max}$ where $\delta n_{\rm max}$ is the maximum refractive-index variation for the used experimental parameters. To be clear, the experimental parameters directly relevant for the study of the hydrodynamics of the fluid of light thus are i) the input angle $\theta$ of the beam which generates it, corresponding to the flow velocity $v$, and ii) the diameter $d$ of the beam producing the optical defect, which controls the effect of the obstacle on the flow.\\
Figure \ref{fig:fig_2} shows typical images of the fluid of light beam intensity distribution (top row) and the corresponding reconstructed phase (bottom row). First, panel (a) corresponds to the superfluid phase. It presents an absence of long-range radiation from the obstacle \cite{amo2009, michel2018}. 
In the other panels, we observe optical vortices with their distinctive behaviour: a deep localised intensity drop associated with a phase singularity. 
To be more precise, one observes in panel (b) a pair of vortices created at the boundary of the obstacle, downstream. They might not be clearly visible in the intensity image, but appear unambiguously in the phase representation. In panels (c) and (d), the same description can be made, both on the intensity and phase distributions, but respectively for two and three pairs of vortices. 
It is worth noting that the vortices always appear in the wake of the obstacle and in pairs with opposite circulations $\oint\mathbf{v}\cdot{\rm d}\mathbf{r}=\pm 2\pi/(\nr\kf)$, where we recall that $\mathbf{v}=\nabla_{\mathbf{r}}\arg(\psif)/(\nr\kf)$ is the local velocity of the fluid. They arise to a certain extent as in classical hydrodynamics where Karman vortex streets may appear in the wake of moving objects. 
This is illustrated by panel (g) which is a zoom of the phase image of panel (d). The region of interest corresponds to the two clear localised intensity drops observed in the top panel (d). The stream plot of the phase derivative, \textit{i.e.}, of the velocity (black lines and arrows), indeed indicates the opposite circulations of the vortices. 
This behaviour is in good qualitative accordance with theoretical and numerical studies reported in the literature (see \cite{engels2010} for a pedagogical viewpoint on the subject). 
In panel (e), one can still picture vortices in the phase, but also an oscillating structure (indicated by the white arrow in the intensity), resulting on the merging of two pairs of vortices, which is often referred as a snake instability in the literature \cite{mironov2010, griffin2020b}. In the last panel (f), this snake instability moves away from the obstacle, and forms likely a dark soliton (indicated by the white arrow). As expected, no phase singularity is observed for both the snake instability and the soliton.
\begin{figure}[b]
\centering
\includegraphics[width=\linewidth]{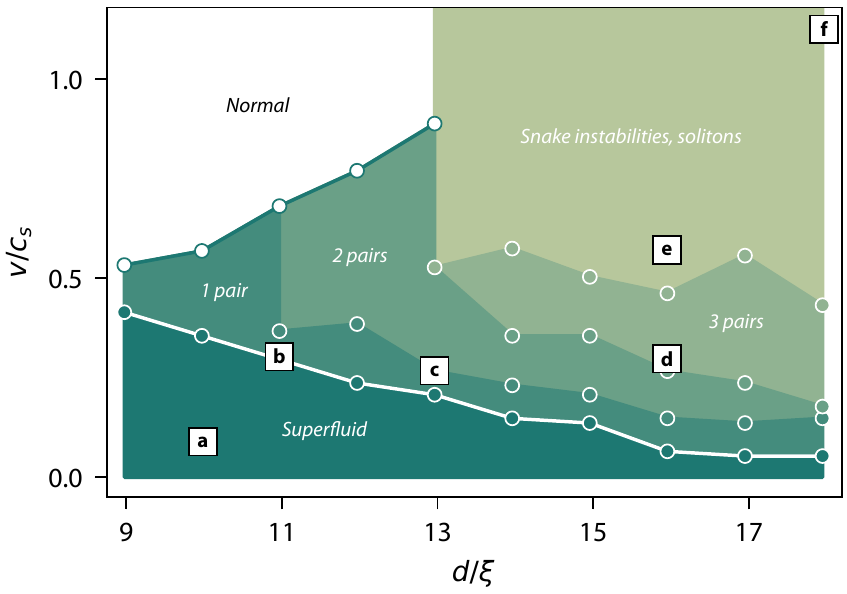}
\caption{Diagram of the observed phases of the fluid of light, ranging from normal (white) to superfluid (dark green). The phases in between (green color scale) correspond to the observation of turbulent coherent structures: vortex pairs, snake instabilities and solitons. Circles are experimental data delimiting the different phases. The squared letters in each phase correspond to the images in fig. \ref{fig:fig_2}.}
\label{fig:fig_3}
\end{figure}

A systematic analysis of such images for various $v/\cs$ and $d/\xi$ allows us to build, in the spirit of refs. \cite{huepe2000, pinsker2014}, a (${d/\xi,v/c_{\rm s}}$) diagram displaying the different hydrodynamic behaviours that the fluid of light exhibits when flowing past the obstacle. This diagram is shown in fig. \ref{fig:fig_3}. Each region is represented by a color (green color scale) and each point corresponds to the apparition of a new distinct feature. For instance, the dark green points, linked by white segments to guide the eye, correspond to the set of flow and obstacle parameters for which one vortex pair suddenly appears by increasing the flow velocity. The squared letters indicate the location on the diagram of the intensity and phase images of fig. \ref{fig:fig_2}. Note however that the precise shape and location of the diagram boundaries depend on the crystal length, which sets the total propagation time of the fluid. Indeed, the turbulent structures observed need a finite time (which depends on the fluid and obstacle parameters) to be created, which means that their number will scale with the ratio of the system size to their respective nucleation time.\\
As expected, this diagram emphasises the existence of three main regimes of transport, namely, a stationary nonsuperfluid, normal regime at high velocity (white), a nonstationary turbulent regime at intermediate velocity, where nonlinear structures such as vortices are abundant (intermediate green colors), and the superfluid regime which is characterised by the absence of long-range excitations (dark green). In addition, few salient features can be commented. \\
First, we clearly observe that as $d/\xi$ increases, the turbulent regime becomes predominant. This regime tends to vanish for low $d/\xi$, as reported in our previous experiment \cite{michel2018} on the normal/superfluid transition triggered by a narrow obstacle ($d\simeq\xi$). By interpolation, the merging point of the upper (normal/turbulent) and lower (turbulent/superfluid) separatrices is estimated to be reached at a nonzero value of $d/\xi$ for a Mach number $v/\cs<1$. This is likely due to photon absorption in the medium (about $1.5~\mathrm{dB}/\mathrm{cm}$), in qualitative agreement with previous theoretical works on low-dimensional dissipative condensates \cite{larre2012}. A full theoretical understanding of these phase boundaries is however still lacking and is currently subjected to further investigation.
Second, for intermediate values of $d/\xi$, from 9 to 13, the turbulent phase mainly consists in the generation of vortex pairs and both normal (white) and superfluid (dark green) phases still exist at respectively high and low flow velocities. 
Then, for larger $d/\xi$, more complex turbulent behaviours appear, involving more vortex pairs, snake instabilities or solitonic structures.
For the value of $v/\cs$ investigated in the experiment, we were not able to recover a stationary nonsuperfluid phase for a wide obstacle. 
Note that the three vertical transitions visible in fig. \ref{fig:fig_3} are actually not as sharp as represented on the diagram. Experimental data are missing in these areas and the analysis of these transition is still under investigation.

The diagram of fig. \ref{fig:fig_3} reveals the rich variety of the turbulent regimes appearing in a 2D fluid of light in a propagating geometry. Although these different behaviours were one by one studied both experimentally and theoretically in ultracold-atom \cite{pitaevskii2003} and cavity-polariton \cite{carusotto13} physics, this is the first time, to the best of our knowledge, that this rich zoology is experimentally explored in its entirety in the context of 2D fluids of light. Here, we gather all the different behaviours in one unique general diagram involving the two main parameters of the constrained flow: the Mach number and the obstacle diameter. It allows us to predict, for a given Mach number and a given obstacle diameter, the nature of the flow (normal/turbulent/superfluid) and the possible turbulent structures that will be supported by the system. Many aspects are still under investigation. In the following, we will focus on a detailed study of the spatial profiles of isolated optical vortices and their dependence on the healing length.

\section{Quantitative analysis of an isolated vortex}

As already discussed, one manifestation of quantum turbulence in 2D fluids of light is the generation of optical vortices characterised by a vanishing intensity associated with a phase singularity. The intensity profile of such a coherent structure is phenomenologically expected to vary over a length scale close to the healing length $\xi$ \cite{isoshima1999, konotop2000, coddington2004}. 
Following ref. \cite{pethick_smith_2008}, the intensity profile of one of our vortices as a function of the radial coordinate $r$ from the core is approximately given by
\begin{equation}
    I_{\rm v}(r)= I_0 \frac{(r/\xi)^2}{1+(r/\xi)^2},
    \label{eq:vortex}
\end{equation}
where $I_0 = I(r\gg\xi)$ is the intensity of the fluid of light far from the vortex core.
Defining the vortex radius $r_{\rm v}$ at $I_{\rm v}(r_{\rm v})/I_0 = s$, where $s$ is an arbitrary parameter smaller than 1, we readily get
\begin{equation}
    r_{\rm v}(s)=\sqrt{\frac{s}{1-s}}\xi.
    \label{eq:rv}
\end{equation}
Equation (\ref{eq:rv}) predicts a linear dependence of the vortex core radius on the healing length, and the proportionality coefficient directly depends on the intensity threshold $s$ where the radius is measured.

\begin{figure}[t!]
\centering
\includegraphics[width=\linewidth]{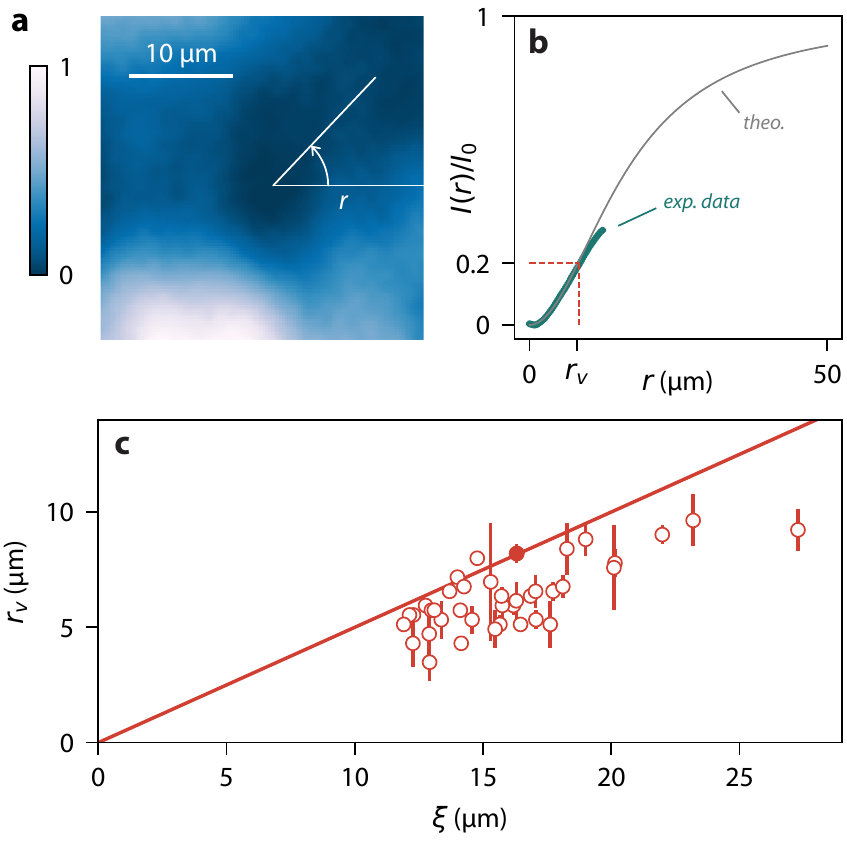}
\caption{(a) Intensity distribution of an isolated vortex normalised to the maximum value of the image. The white bar corresponds to 10 $\mu$m. 
(b) Intensity radial profile $I(r)$ obtained after an azimuthal integration (see (a)). The profile is normalised by the intensity $I_0$ measured in the absence of any obstacle. The green curve corresponds to the experimental data. The dashed gray curve is the approximate theoretical intensity profile $I_{\rm v}(r)$ obtained for $\xi = 16.3$ $\mu$m. The vortex radius $r_{\rm v}$ is measured at the threshold $s = I(r)/I_0 = 0.2$. The corresponding value is reported in (c) (filled red circle). 
(c) Vortex radius versus $\xi$. 
The red circles correspond to experimental data. The error bars are the standard deviations of the vortex radii measured on the same image.
The red line corresponds to the theoretical law, $r_{\rm v} = 0.5 \xi$ obtained for $s= 0.2$.}
\label{fig:fig_4}
\end{figure}

To experimentally investigate the vortex size as a function of the healing length in a fluid of light, we track and study them under various conditions. More precisely, the fluid of light intensity is varied from 6 to 16 mW/cm$^2$  to get different values of $\xi$, ranging from 15 to 30 $\mu$m. We performed different sets of measurements, for different values of the obstacle diameter and strength. As detailed in the Supplementary Material, keeping a constant strength for different values of the fluid of light intensity imposes to vary the obstacle beam intensity as well. Here it ranges from 1.5 to 4 W/cm$^2$, which corresponds to refractive-index variations from 1.17 to 2.09$\ttt{-4}$ thus ranging from 0.7 to 0.8 $\delta n_{\rm max}$. The obstacle diameter varies from 60 to 80 $\mu$m and the fluid of light velocity is fixed to 2$\ttt{-3}$.\\
As seen in the previous section both intensity and phase are measured.
To extract the vortex sizes from each images, we perform the following processing. First, the vortex is isolated from its environment (\textit{e.g.}, a neighbouring vortex or any other intensity fluctuation). To do so, the position of the vortex core is tracked from the phase singularity and the intensity image is cropped to 25 $\mu$m $\times$ 25 $\mu$m (see fig. \ref{fig:fig_4}(a)). Second, we perform a $2\pi$ azimuthal integration of the intensity profile. The profile is normalised to the unperturbed intensity $I_0$ in the absence of obstacle. The corresponding dimensionless radial profile $I(r)/I_0$ is plotted in fig. \ref{fig:fig_4}(b) (green data). Then, the radius of the vortex $r_{\rm v}$ is measured at $s = I(r)/I_0 = 0.2$ (red dashed horizontal and vertical lines in fig. \ref{fig:fig_4}(b)). Above this threshold, the profiles of vortices with the highest radii are drastically affected by the environment. The theoretical prediction is represented by the gray dashed line, calculated from eq.~\eqref{eq:vortex} with $\xi$ = 16.3 $\mu$m, the measured experimental value for the vortex pictured in fig. \ref{fig:fig_4}(a).\\
All the measured vortex radii are plotted in fig. \ref{fig:fig_4}(c) as a function of $\xi$. The error bars are the standard deviations of the vortex radii measured on the same image. The filled red circle highlights the measure of the vortex radius depicted in fig. \ref{fig:fig_4}(a) and we used the corresponding healing length to plot the theoretical prediction in fig. \ref{fig:fig_4}(b).
For comparison, the prediction, $r_{\rm v}=0.5 \xi$ for $s=0.2$ in eq. \eqref{eq:rv}, is plotted (solid red line).
Interestingly, the experimental data are in rather good agreement with this theoretical expectation.
Plotting separately the data corresponding to a given experimental configuration (\textit{e.g.}, fixed obstacle strength and diameter) does not exhibit some specific tendency. All experimental points converge to the same behaviour, meaning that, as discussed above, the vortex size only depends on the main characteristic length scale of the nonlinear system, namely, the healing length. Neither the environment (\textit{e.g.}, the obstacle diameter and strength) nor the initial fluid velocity seem to strongly influence the size of the vortex core.

\section{Conclusion}

In this letter, we have presented an experimental study of the breakdown of superfluidity and of the various regimes of transport in a 2D fluid of light in a propagating geometry.
First, we have experimentally built a phase diagram gathering the different types of flow and coherent nonlinear excitations characterising them, in between the stationary normal and superfluid phases at high and low velocity, respectively. 
Second, we have proposed a quantitative study of the radii of the vortices nucleated downstream from the obstacle, and showed a good qualitative agreement with theoretical estimates.
Our experimental platform gives the opportunity to study the rich nonlinear hydrodynamics of a superfluid of light flowing past a single optical defect. However, there is no major experimental restriction to design more complex environments made of randomly distributed obstacles, opening the way to the study of the transport of superfluid light in disordered landscapes.

\acknowledgments
The authors thank S. \textsc{Nazarenko} and S. \textsc{Rica} for fruitful discussions.
This work received funding from the European Union Horizon 2020 research and innovation program under Grant Agreement No. 820392 (PhoQuS), the French government, through the UCA$^\textrm{JEDI}$ Investments in the Future project managed by the National Research Agency (ANR) with the reference number ANR-15-IDEX-01 and the Region Sud.

\bibliographystyle{naturemag}
\bibliography{bibliography}

\end{document}


\title{Supplementary materials: Experimental observation of turbulent coherent structures in a superfluid of light}

\author{A. Eloy}
\affiliation{Universit\'e C\^ote d'Azur, CNRS, INPHYNI, France}
\author{O. Boughdad}
\affiliation{Universit\'e C\^ote d'Azur, CNRS, INPHYNI, France}
\author{M. Albert}
\affiliation{Universit\'e C\^ote d'Azur, CNRS, INPHYNI, France}
\author{P.-\'E. Larr\'e}
\affiliation{Universit\'e C\^ote d'Azur, CNRS, INPHYNI, France}
\author{F. Mortessagne}
\affiliation{Universit\'e C\^ote d'Azur, CNRS, INPHYNI, France}
\author{M. Bellec} 
\affiliation{Universit\'e C\^ote d'Azur, CNRS, INPHYNI, France}
\author{C. Michel}
\affiliation{Universit\'e C\^ote d'Azur, CNRS, INPHYNI, France}

\maketitle

\section{Photorefractive effect and effective nonlinear index variation}

The basic mechanism of the photorefractive (PR) effect remains in the photogeneration and displacement of mobile charge carriers driven by an external electric field $E_0$ \cite{denz2003}.
The induced permanent space-charge electric field thus implies a modulation of the refractive index of the crystal, 
\begin{equation}
    \Delta n(I, \mathbf{r}) = -0.5\nr^3r_{33}E_0\frac{I(\mathbf{r})/I_{\rm sat}}{1 + I(\mathbf{r})/I_{\rm sat}},    
\end{equation}
where $\nr$ is the optical refractive index and $r_{33}$ the electro-optic coefficient of the material along the extraordinary axis, $I(\mathbf{r})$ is the intensity of the optical beam in the transverse plane, and $I_{\rm sat}$ is the saturation intensity which can be adjusted with a white light illumination of the crystal, and $E_0$ is an external electric field applied to the crystal along the $c$-axis of the crystal.
For a crystal of strontium-barium-niobate (SBN:61), $r_{33}=235$ pm/V and $\nr=2.36$.
The experimental control over $E_0$ and $I_{\rm sat}$ allows to precisely tune $\Delta n(I, \mathbf{r})$ \cite{boughdad2019}.

In the main paper, we write the propagation eq. (1), formulated to mimic the Gross-Pitaevskii-like equation, with a nonlinear term, $\kf\Delta n(\If)$, representing the interactions and an external potential, $\kf\delta n(\Iob)$ acting as a potential barrier.
However, there is a more realistic manner to model the nonlinear interactions and the local depletion of the refractive index. 
We have to consider the propagation of two coupled beams, the one acting as the fluid of light, and the one acting as the obstacle, in a PR crystal considering a saturable \textsl{isotropic} nonlinearity. 
The propagation equations thus read:
\begin{eqnarray}
	i\partial_z\psi_{\rm f} &=& -\frac	{1}{2k_{\rm f}n_{\rm e}}\nabla^2_{\bf r} \psi_{\rm f} - k_{\rm f}\Delta n(I_{\rm f}+I_{\rm ob})\psi_{\rm f}\\
	i\partial_z\psi_{\rm ob} &=& -\frac	{1}{2k_{\rm ob}n_{\rm e}}\nabla^2_{\bf r} \psi_{\rm ob} - k_{\rm ob}\Delta n(I_{\rm f}+I_{\rm ob})\psi_{\rm ob}
\end{eqnarray}
where $\psi_{\rm f}$ and $\psi_{\rm ob}$ are the slowly varying envelopes of the optical fields for the fluid and the obstacle, respectively, and considering the bulk refractive index does not change with the wavelength.
The coupling thus comes from the nonlinear term.
Indeed, the total nonlinear refractive index of the medium varies with the two intensities $I_{\rm f,ob}=|\psi_{\rm f,ob}|^2$ as
\begin{equation}
	\Dntot=\Delta n(I_{\rm f}+ I_{\rm ob})=-0.5\nr^3r_{33}E_0\frac{\tilde I_{\rm f}+\tilde I_{\rm ob}}{1+\tilde I_{\rm f}+\tilde I_{\rm ob}}
	\label{eq:dntot}
\end{equation}
with $\nr=2.36$ and $r_{33}=235$ pm/V for a SBN:61 crystal with a linear polarisation along the polar axis (c-axis) of the crystal, $E_0$ the external voltage applied to the crystal, and $\tilde I_{\rm f,ob}=I_{\rm f,ob}/I_{\rm sat}$ the laser intensities normalised to the saturation intensity $I_{\rm sat}$. $\Delta n(I)$ saturates for intensities much higher that $I_{\rm sat}$ and the absolute maximum value is $|\Delta n_{\rm max}|=0.5n_0^3r_{33}E_0$. 
With the typical experimental values $E_0=1300$ V/cm and $\Isat=400$ mW/cm$^2$, we have $\Delta n_{\rm max}=-2.0\times 10^{-4}$.

\begin{figure}
    \centering
	\includegraphics[width=0.85\columnwidth]{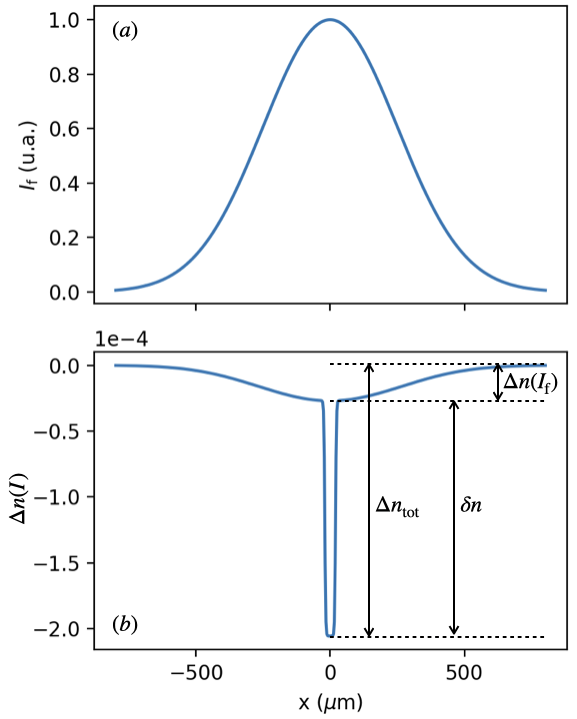}
	\caption{(a) Gaussian initial condition for the fluid beam, (b) effective index variation for the fluid beam, when it modifies itself the refractive index of the material through the photorefractive effect.}
	\label{fig:supp-fig-1}
\end{figure}

Figure \ref{fig:supp-fig-1}(a) illustrates the input Gaussian beam, corresponding to the initial intensity profile for the fluid of light. The latter propagates in a medium in which we impose a local index variation $\Delta n(I_{\rm ob})=\delta n=-1.5\ttt{-4}$. Figure \ref{fig:supp-fig-1}(b) represents the total index variation, $\Delta n_{\rm tot}=\delta n + \Delta n(I_{\rm f})$, reflecting the fact that the fluid of light also contributes, nonlinearly, to the refractive index variation. As a consequence, the maximum variation of the total index of refraction is greater than $\delta n$, but in the example, this does not influence the fluid of light.
However, a problem appears when the intensity of the obstacle beam makes the nonlinear refractive index variation saturate, which is commonly the case in this kind of experiments. Indeed, in this case, the nonlinear contribution imposed by the fluid itself induces an effective refractive index, $\Delta n_{\rm eff}$, lower than the maximum refractive index, $\Delta n_{\rm max}$, fixed by the obstacle.
In order to evaluate the effective refractive index variation felt by the fluid of light, we develop an extension of the usual formula for the nonlinear index variation in a photorefractive crystal of SBN:61 \cite{denz2003}.
We consider that the effective $\Delta n_{\rm eff}$ experienced by the fluid can be written as
\begin{equation}
	\Delta n_{\rm eff} = \Delta n(I_{\rm f}+ I_{\rm ob})-\Delta n(I_{\rm f}).
	\label{eq:dneff_tot}
\end{equation}

\medskip
When assuming that the obstacle beam makes the nonlinear index saturate: $\Delta n(\Iob)=\Delta n_{\rm max}$, the previous equation reads
\begin{eqnarray}
	\Delta n_{\rm eff} &=& \Delta n_{\rm max}-\Delta n(I_{\rm f})\nonumber\\
	&=& \Delta n_{\rm max}-\Delta n_{\rm max}\frac{\tilde I_{\rm f}}{1+\tilde I_{\rm f}}\nonumber\\
	&=& \Delta n_{\rm max} \frac{1}{1+\tilde I_{\rm f}}.
	\label{eq:dneff_If}
\end{eqnarray}
It is then obvious that the effective refractive index seen by the fluid of light decreases as its intensity increases.

\begin{figure}
    \centering
    \includegraphics[width=0.95\columnwidth]{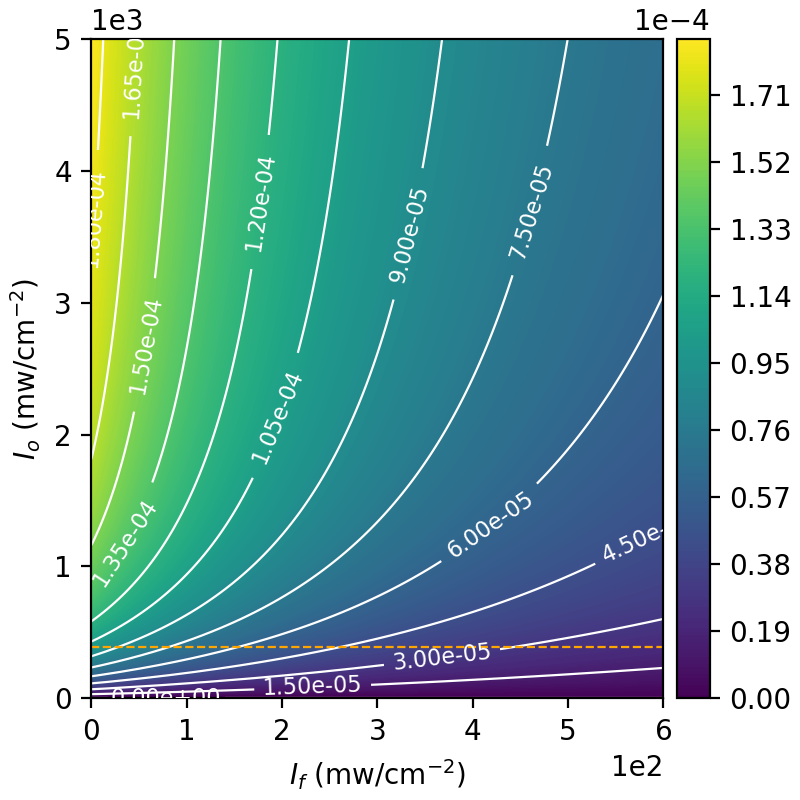}
    \caption{Cartography of the absolute value of the effective nonlinear index variation experienced by the fluid as a function of the fluid intensity on the horizontal axis, and the obstacle intensity on the vertical axis. The saturation intensity is denoted by the orange dashed horizontal line. The white lines correspond to iso-index lines.}
    \label{fig:supp-fig-3}
\end{figure}

\medskip
If we take the complete expression of eq. (\ref{eq:dneff_tot}), then we need to plot a map of $\Delta n_{\rm eff}$ depending on $I_{\rm f}$ as well as on $I_{\rm ob}$. To do so, one can develop the expression of $\Delta n_{\rm eff}$ such as
\begin{equation}
	\Delta n_{\rm eff} = \Delta n_{\rm max}\left[ \frac{I_{\rm ob}/I_{\rm sat}}{1+\frac{2I_{\rm f}}{I_{\rm sat}}+\frac{I_{\rm ob}}{I_{\rm sat}}+\frac{\left(I_{\rm f}+I_{\rm ob}\right)I_{\rm f}}{I_{\rm sat}^2}} \right],
	\label{eq:dneff_dev}
\end{equation}
whose absolute value is plotted as a parametric plot in fig. \ref{fig:supp-fig-3}. In this figure, we represent a contour plot whose different colors correspond to different values of $|\Delta n_{\rm eff}|$, with $I_{\rm ob}$ on the vertical axis and $I_{\rm f}$ on the horizontal axis. The white lines correspond to iso-index lines, and the orange dashed horizontal line corresponds to a typical $I_{\rm sat}$ of 400 mW/cm$^2$.

Inspecting the limits of this expression, we can verify that we have the right intuition:
\begin{itemize}
	\item if $I_{\rm ob}\gg I_{\rm sat} \gg \If$, then $\Delta n_{\rm eff}\rightarrow\Delta n_{\rm max}$, which is expected, as in this case, the obstacle beam completely makes the nonlinear index saturate. This corresponds to the upper left corner of the figure.
	\item if $I_{\rm ob}\ll I_{\rm f}$ and $I_{\rm f} \gg I_{\rm sat}$, then $\Delta n_{\rm eff}\rightarrow 0$. This is also expected as in this condition, the effect of the obstacle is overshadowed by the fluid, and if the fluid saturates the medium, the system goes back to linear. This corresponds to the lower right corner of the figure.
\end{itemize}

In the intermediate cases, it becomes obvious that the fluid beam almost never sees the maximum value of the nonlinear index variation. However, one can manage to optimise this value, and, more important, to keep it constant while varying $I_{\rm ob}$ and $I_{\rm f}$, following the white lines of fig. \ref{fig:supp-fig-3}. This is what is done in the experiments presented in the main paper.

\section{Bogoliubov relation dispersion for a saturable nonlinearity}

\begin{figure}
    \centering
    \includegraphics[width=\columnwidth]{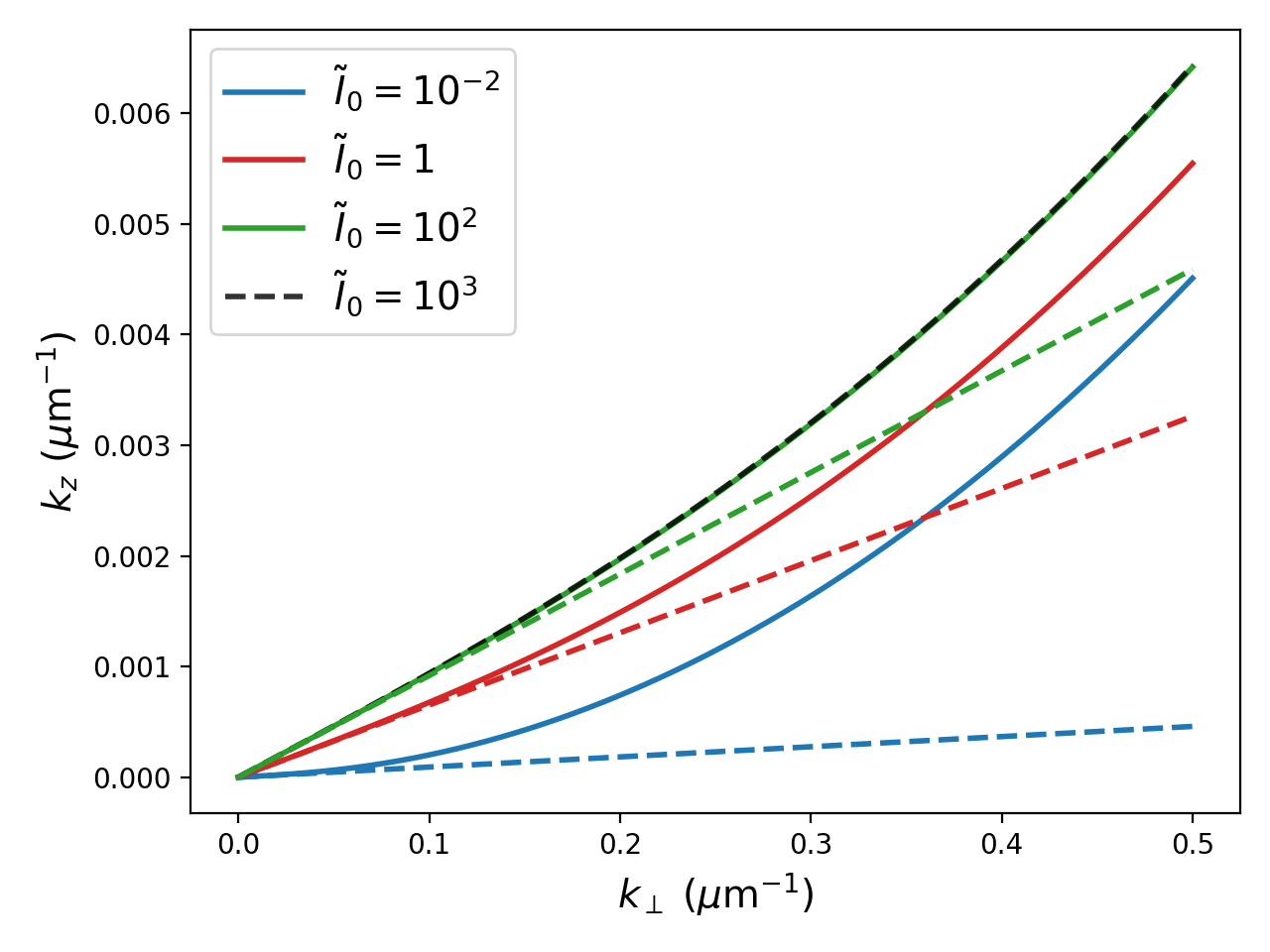}
    \caption{Dispersion relation of the elementary excitations on top of a homogeneous 2D fluid of light at rest. Blue, red and green solid lines, as well as the dotted-black line correspond to the dispersion relation (\ref{eq:bog}) for different values of $\tilde I_0$. The dashed lines correspond to the $k_\perp\ll1/\xi$ sonic regime, dominated by the nonlinearity. The slope of the straight lines directly provides the Bogoliubov sound velocity. In the deep saturation regime, the Bogoliubov dispersion relation is independent of the value of $\tilde I_0$, as the superposition of the green and black-dotted curves reveals.}
    \label{fig:supp-fig-4}
\end{figure}

In this section, we adapt Bogoliubov's theory of linearised fluctuations to get the dispersion relation of the small modulations of the fluid of light in the nonlinear photorefractive crystal, in the simplest configuration where the fluid is uniform and at rest in the transverse plane. To do so, we start from the following nonlinear Schrödinger equation for the slowly varying wavefunction of the fluid:
\begin{equation}
    i\partial_z\psi =-\frac{1}{2k \nr}\nabla^2\psi - k\Delta n(I)\psi
\end{equation}
where $\Delta n(I)=\Delta n_{\rm max}\frac{\tilde I}{1+\tilde I}$, with $\tilde I=I/\Isat$ and $I=|\psi|^2$. Using the Madelung transformation $\psi(\mathbf{r})=\sqrt{I(\mathbf{r})} e^{i S(\mathbf{r})}$, it comes
\begin{eqnarray}
    \partial_z I+\frac{1}{k\nr}\bnabla(I\bnabla S)=0,\\
    \partial_z S = \frac{1}{2k\nr}\frac{\nabla^2\sqrt{I}}{\sqrt{I}}-\frac{1}{2k\nr}\left(\bnabla S\right)^2+k\Delta n(I),
\end{eqnarray}
Considering small perturbations on top of the homogeneous fluid at rest:
\begin{eqnarray}
    I&=&I_0+\delta I,\\
    S&=&k\Delta n(I_0)z+\delta S,
\end{eqnarray}
leads to the following equations for the perturbations:
\begin{eqnarray}
    \partial_z\delta I &=& -\frac{1}{k\nr}I_0\nabla^2\delta S,\\
    \partial_z \delta S &=& \frac{1}{4k\nr}\frac{1}{I_0}\nabla^2\delta I+k\Delta n_{\rm max}\frac{\delta\tilde I}{(1+I_0^2)}.
\end{eqnarray}
Deriving the latter with respect to $z$ and combining it with the former, we eventually get, for the phase fluctuations,
\begin{equation}
     \partial_z^2\delta S=-\frac{1}{4k^2\nr^2}\nabla^4\delta S + \frac{\Delta n(I_0)}{\nr}\nabla^2\delta S.
\end{equation}
To get the dispersion relation of the fluctuations on top of the fluid, we search for the solution of eq. (16) in the plane-wave form $S\propto\exp{\left[i\left( \mathbf{k}_\perp\cdot\mathbf{r}+k_z z \right)\right]}$, which leads to the following Bogoliubov-type dispersion relation:
\begin{equation}
    k_z=\sqrt{\frac{k_\perp^2}{2k\nr}\left( \frac{k_\perp^2}{2k\nr} +2k\Delta n(I_0)\right)}.
    \label{eq:bog}
\end{equation}

From this equation, one extract the healing length and the sound velocity, which are
\begin{eqnarray}
    \xi &=& \frac{1}{\sqrt{k^2\nr|\Delta n(I_0)|}},\\
    \cs &=& \sqrt{\frac{|\Delta n(I_0)|}{\nr}}.
\end{eqnarray}

Figure \ref{fig:supp-fig-4} displays the Bogoliubov dispersion relation as a function of the transverse wavenumber $k_\perp$ for different values of $\tilde I_0$ and typical experimental parameters. Obviously, when $\tilde I_0\ll 1$, the nonlinear contribution is negligible, and the sound velocity, corresponding to the slope of the linear part of the dispersion relation at low $k_\perp$ tends to $0$. On the other hand, when $\tilde I_0\gg 1$, the saturation manifests in the fact that the shape of the dispersion relation does not change anymore when increasing $\tilde I_0$. However, the typical linear and parabolic trends of the dispersion for respectively $k_\perp\ll 1/\xi$ and $k_\perp\gg 1/\xi$ remain. Nevertheless, it is important in the experiment that the hydrodynamical parameters (\textit{i.e.}, $\cs$, $\xi$) still depend on the fluid density, so we take care of never exceeding $\tilde I_0=1$.

\bibliographystyle{plain}
\bibliography{bibliography}